# Self-Oscillating Photonic Frequency Comb for Sub-THz Local Oscillator Generation in Earth Observation Systems

Dimitrios Kastritsis*[a], Ioannis Stratakis[a], Stavros Iezekiel[a]
[a] Emphasis Research Centre, University of Cyprus, Nicosia, Cyprus

## ABSTRACT

Earth observation missions rely on microwave and sub-terahertz radiometry for retrieving atmospheric parameters such as temperature, humidity, and cloud properties, with the 183- and 229-GHz spectral regions being particularly relevant. Local-oscillator (LO) generation at these frequencies requires low phase noise and high frequency stability while meeting stringent size, weight, and power (SWaP) constraints. Conventional architectures employ dielectric resonator oscillators followed by cascaded frequency-multiplication stages, requiring dedicated generation chains for different frequency bands and increasing hardware complexity. Furthermore, frequency multiplication degrades the phase noise of the generated signal. This work proposes and numerically investigates a reconfigurable self-oscillating microwave photonic architecture for LO generation in Earth observation receivers. Cascaded Mach–Zehnder and phase modulation within an optoelectronic feedback loop generate an optical frequency comb with enhanced higher-order components. A combination of optical filters is used to select the required comb line. First, a conventional bandpass filter with a bandwidth of approximately 10 GHz isolates the spectral region containing the desired component, after which a narrowband dual cavity Bragg grating ring resonator (BGRR) with a bandwidth of approximately 3 MHz provides the final line selection. The selected comb line is then combined with an optical reference derived from the same laser and photodetected to generate the desired millimeter-wave or sub-terahertz signal through optical heterodyning. Numerical simulations are presented for 183- and 229-GHz generation, demonstrating the ability of the same underlying architecture to address multiple LO frequencies by tuning the OEO electrical filter to the appropriate oscillation frequency and selecting the corresponding comb order.

**Keywords:** Self-oscillating frequency comb, optoelectronic oscillator, microwave photonics, Earth observation

*Email: kastritsis.dimitrios@ucy.ac.cy

## 1. INTRODUCTION

Millimeter-wave and sub-terahertz technologies play a key role in satellite-based Earth observation, particularly in microwave sounding for atmospheric temperature and humidity profiling, as exemplified by the MetOp Second Generation (MetOp-SG) Microwave Sounder (MWS), whose 24 channels span frequencies from 23.8 to 229 GHz [1]. Within this frequency range, the 183-GHz band (175.311–191.311 GHz), used for atmospheric water-vapor profiling, and the 229-GHz quasi-window (228–230 GHz) are of particular relevance to the present work [1], [2].

Spectrally pure local-oscillator (LO) generation is a key requirement at these frequencies, with the MWS relying on cascaded electronic frequency-multiplication chains to generate the required LO signals. Specifically, the 183-GHz receiver employs 30.552-GHz dielectric resonator oscillators (DROs) with integrated active doublers, followed by Schottky tripling, whereas the 229-GHz receiver employs 28.625-GHz DROs together with passive MMIC and Schottky frequency-doubling stages [2]. Although well established, such electronic LO architectures rely on multiple frequency-multiplication stages to reach the required operating frequencies, motivating the investigation of alternative approaches offering greater flexibility and reconfigurability.

Microwave photonics provides an attractive approach to high-frequency signal generation by exploiting the large bandwidth and low propagation loss available in the optical domain [3]. In particular, optoelectronic oscillators (OEOs) employ optical energy storage within a feedback loop to achieve high effective quality factors and generate spectrally pure microwave and millimeter-wave signals with low phase noise [4], [5]. For Earth-observation applications, a 45.86-GHz OEO investigated as a potential intermediate LO source for the MWS, Microwave Imager (MWI), and Ice Cloud Imager (ICI) demonstrated a sidemode suppression of approximately 46 dB and a phase noise of −102 dBc/Hz at a 100-kHz offset [6].

Optical frequency combs can further extend photonic signal generation toward millimeter-wave and sub-terahertz frequencies through photodetection of appropriately spaced optical spectral components [7] - [9]. Electro-optic frequency-comb generation is particularly attractive for microwave-photonic applications because the comb spacing is directly related to the frequency of the electrical signal driving the electro-optic modulators, while cascaded modulation provides flexible control over the number and relative amplitudes of the generated comb lines. Such approaches have been extensively investigated using intensity, phase, and cascaded electro-optic modulators [8], [9]; however, conventional electro-optic comb generators generally rely on an external microwave source to drive the modulators.

Self-oscillating optical frequency-comb generators (SOFCGs) combine electro-optic comb generation with optoelectronic feedback, allowing the electrical drive signal required for modulation to be generated within the feedback loop itself [7], [8]. A dual-drive Mach–Zehnder modulator (MZM)-based SOFCG demonstrated 23 comb lines with an 11.84-GHz spacing and the generation of a 94.8-GHz signal by optical heterodyning, with potential extension to approximately 260 GHz [7]. An alternative architecture employing cascaded MZM and phase modulation generated eleven comb lines with 10- and 12-GHz spacings, with the spectral power distribution controlled through the modulation amplitudes, relative phase, and MZM bias [8]. These results demonstrate the potential of self-oscillating comb architectures to combine microwave oscillation, optical-comb generation, and high-frequency signal synthesis within a common photonic platform.

Building on these concepts, this work investigates self-oscillating photonic architecture for the reconfigurable generation of 183- and 229-GHz LO signals for Earth-observation applications. The proposed architecture combines cascaded Mach–Zehnder and phase modulation, an optoelectronic oscillator (OEO) feedback loop, and narrowband tunable optical filtering within a common photonic–electronic platform. By appropriately selecting the OEO oscillation frequency and tuning the optical filter to the corresponding comb component, different LO frequencies can be addressed using the same generation architecture, potentially reducing the need for dedicated oscillator and frequency-multiplication chains. Numerical simulations are presented for 183- and 229-GHz generation to demonstrate the reconfigurability of the proposed approach. Suitable optical functions may also be amenable to implementation using photonic integrated technologies.

The remainder of this paper is organized as follows. Section II presents the proposed architecture and its operating principle, while Section III describes the numerical model and presents the results for 183- and 229-GHz generation. Finally, Section IV summarizes the main conclusions.

## 2. PRINCIPLE OF OPERATION AND FREQUENCY GENERATION CONCEPT

The block diagram of the proposed architecture is illustrated in Fig. 1. The self-oscillating frequency-comb combines an optoelectronic oscillation loop with an optical filtering and heterodyne-generation section. A continuous-wave (CW) laser is first split into two optical paths. In the main path, the optical carrier is modulated by the cascaded MZM and phase modulator (PM), generating the optical frequency comb whose line spacing is determined by the OEO oscillation frequency. Since the modulators are driven only at the microwave comb-spacing frequency rather than at the final sub-THz LO frequency, the required electro-optic modulation bandwidth can remain substantially below the final sub-THz LO frequency [9]. The comb signal is amplified by a semiconductor optical amplifier (SOA) and split into two paths by a multimode interference (MMI) coupler. A portion of the optical signal is fed back to a photodetector to close the optoelectronic loop. Following photodetection, the electrical signal is amplified by a low-noise amplifier (LNA) and a power amplifier (PA), spectrally constrained by an electrical band-pass filter (EBPF). An electrical phase shifter adjusts the relative phase of the RF signals driving the MZM and PM before the signal is applied to the PM. Oscillation is established when the loop-gain and loop-phase conditions are simultaneously satisfied. The EBPF restricts the frequency range over which oscillation can occur, while the total loop delay determines the supported longitudinal modes. The resulting oscillation frequency, $f_{\mathrm{OEO}}$, determines the spacing of the generated optical frequency comb.

The remaining portion of the generated comb is first passed through a 10 GHz coarse optical band-pass filter (Coarse OBPF) to suppress out-of-band optical components and provide preliminary spectral selection. A narrowband 3 MHz wavelength-selective filter (WSF) subsequently isolates the desired N-th-order comb line. In parallel, the second optical path provides a copy of the original optical carrier, whose phase can be adjusted before recombination with the selected comb component. The two optical signals are then combined, amplified by an SOA, and applied to a high-frequency photodetector. Their heterodyne beating generates the desired millimeter-wave or sub-THz LO signal at $f_{\mathrm{LO}} = N f_{\mathrm{OEO}}$.

The components highlighted in light blue in Fig. 1 correspond to the photonic subsystem designed for fabrication on an InP platform through the JePPIX multi-project wafer process. This subsystem represents a partial implementation of the proposed architecture and does not include all functions required for complete sub-THz LO generation. In particular, the wavelength-selective optical filtering stage is considered as an external component in the present system model. The results presented in this work are therefore based on numerical system-level analysis rather than experimental validation of complete architecture.

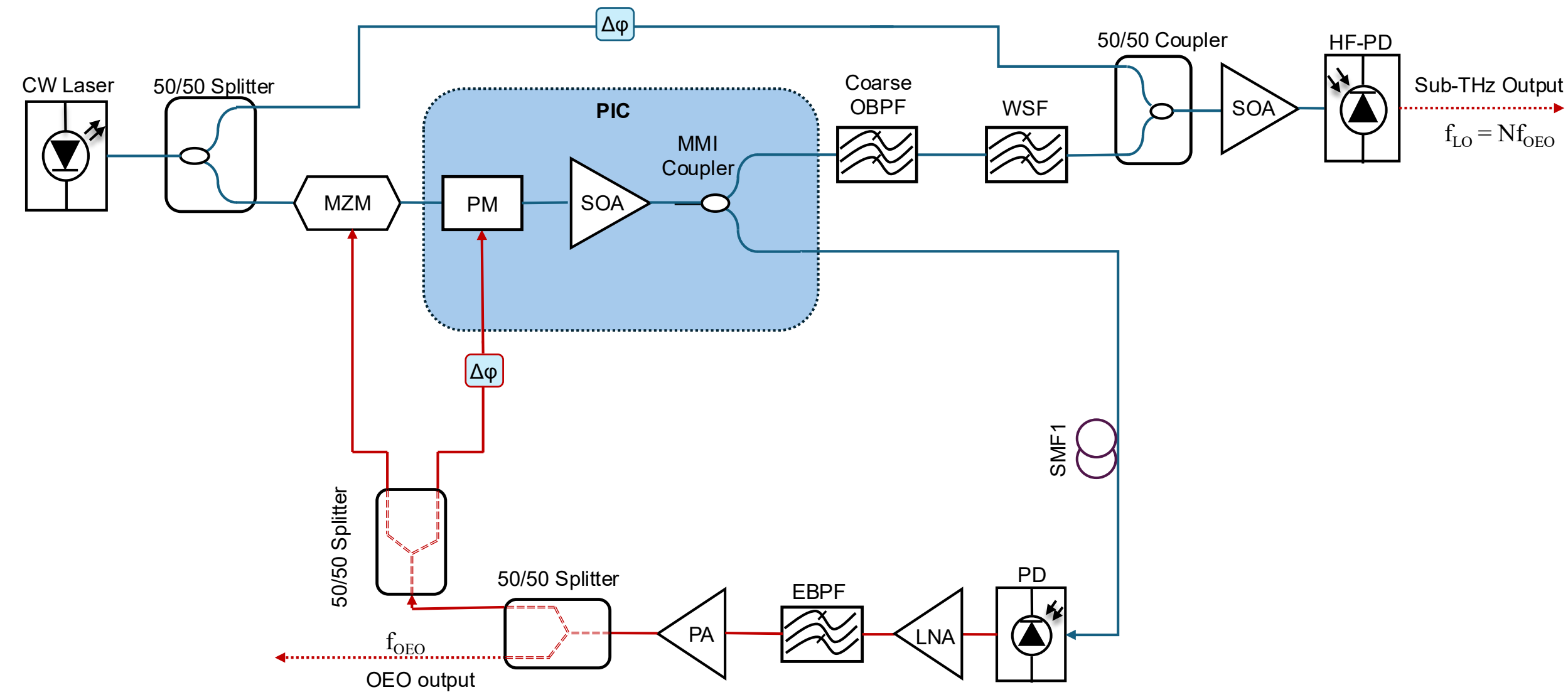

Figure 1. Principle of operation of the proposed self-oscillating photonic frequency-comb architecture for sub-THz LO generation. The OEO defines the optical comb spacing, while the selected N-th-order comb component is recombined with the optical carrier to generate the desired LO frequency through photodetection.

For a required local-oscillator frequency $f_{LO}$, the OEO should operate at an integer subharmonic according to $f_{OEO} = f_{LO}/N$.

Appropriate selection of N allows different required LO frequencies to be generated while keeping the OEO operating frequency within a comparatively narrow microwave-frequency range.

Table 1 summarizes the candidate OEO operating frequencies obtained for the considered MWI, MWS and ICI LO frequencies and different integer multiplication orders.

Table 1. Candidate OEO operating frequencies, $f_{OEO} = f_{LO}/N$, for integer multiplication orders and the considered MWI, MWS and ICI LO frequencies.

| **LO frequency (GHz)** | **N=2** | **N=3** | **N=4** | **N=5** | **N=6** | **N=7** | **N=8** | **N=9** | **N=10** | **N=11** |
|---|---|---|---|---|---|---|---|---|---|---|
| **89** | **44.500** | **29.667** | 22.250 | 17.800 | 14.833 | 12.714 | 11.125 | 9.889 | 8.900 | 8.091 |
| **118** | 59.000 | **39.333** | **29.500** | 23.600 | 19.667 | 16.857 | 14.750 | 13.111 | 11.800 | 10.727 |
| **166** | 83.000 | 55.333 | **41.500** | **33.200** | 27.667 | 23.714 | 20.750 | 18.444 | 16.600 | 15.091 |
| **183** | 91.500 | 61.000 | **45.750** | 36.600 | **30.500** | 26.143 | 22.875 | 20.333 | 18.300 | 16.636 |
| **229** | 114.500 | 76.333 | 57.250 | **45.800** | 38.167 | **32.714** | 28.625 | 25.444 | 22.900 | 20.818 |
| **243** | 121.500 | 81.000 | 60.750 | 48.600 | **40.500** | 34.714 | **30.375** | 27.000 | 24.300 | 22.091 |
| **325** | 162.500 | 108.333 | 81.250 | 65.000 | 54.167 | **46.429** | **40.625** | 36.111 | 32.500 | **29.545** |
| **448** | 224.000 | 149.333 | 112.000 | 89.600 | 74.667 | 64.000 | 56.000 | 49.778 | **44.800** | **40.727** |

The entries highlighted in bold correspond to operating frequencies clustered around approximately 30 GHz and 45 GHz. These two frequency regions are of particular interest because they indicate that several of the required Earth-observation LO frequencies could, in principle, be addressed using an OEO architecture incorporating an electrical band-pass filter with a limited tuning range, rather than requiring a different oscillator chain for every LO band. The approximately 30 GHz operating region was selected for the system-level simulations because it is more compatible with the bandwidth limitations of currently available photonic integrated components, particularly those accessible through the JePPIX platform. The relevance of the higher-frequency operating region is further supported by our previous investigation of OEO-based signal generation in the V-band [6], which showed the feasibility of extending the architecture to substantially higher microwave frequencies. The 45 GHz region is therefore retained in Table 1 to illustrate the broader scalability of the proposed concept, although it is not considered in the present numerical implementation.

In the numerical analysis presented here, the proposed approach is demonstrated for representative LO frequencies of 183 GHz and 229 GHz, corresponding to OEO operation around 30 GHz and selection of the appropriate higher-order comb component

**Wavelength Selective Filter**

The narrowband optical filtering function employed in the proposed self-oscillating optical frequency-comb generator is based on the ultra-narrowband dual-cavity Bragg grating ring resonator (BGRR) previously introduced in [10], [11] and shown schematically in Fig. 2. The structure consists of two cascaded Fabry–Perot (FP) cavities arranged in a closed-loop configuration, with each cavity forming one half of the ring and Bragg gratings acting as the cavity mirrors. The resulting dual-cavity BGRR therefore combines the spectral selectivity of Bragg-grating FP cavities with the resonant properties of a ring resonator. The filter was originally developed for a silica single-mode-fiber (SMF) implementation and modeled using the transfer matrix method (TMM).

In the present work, the structure is extended to provide thermo-optic tunability. In contrast to the previously reported configuration, the refractive indices of the fiber sections forming Cavity 1 and Cavity 2 are varied, while the refractive-index profiles of the Bragg mirrors are kept unchanged. Changing the cavity refractive index modifies its optical path length and consequently shifts the narrow FP resonance within the Bragg stopband. This approach is analogous to previously demonstrated thermally tunable fiber FP filters in which heating is localized predominantly within the cavity region, thereby limiting the thermal perturbation of the surrounding FBG reflectors [12]. For silica fibers, thermal tuning results predominantly from the temperature dependence of the refractive index, although thermal expansion provides an additional, smaller contribution [13].

Accordingly, the thermal tuning in the present numerical model is represented through the cavity refractive index $n_{cav} = n_0 + \Delta n$, where $n_0$=1.44 is the nominal refractive index adopted for silica at 1550 nm and

$$10^{-6} \leq \Delta n \leq 5 \times 10^{-4}.$$

This interval is consistent with realistic thermo-optic variations of silica. Reported values of the thermo-optic coefficient near 1550 nm are of the order of $10^{-5}$ $K^{-1}$; consequently, the considered range corresponds approximately to temperature variations between 0.1 K and 50 K. Thermal expansion is neglected in the present model, since the thermal expansion coefficient of fused silica is more than an order of magnitude smaller than its thermo-optic coefficient.

The Bragg mirrors are implemented as quarter-wave stacks designed for the filter center wavelength $\lambda_0$, with alternating sections characterized by refractive indices $n_1$ and $n_2$ and corresponding physical lengths $d_1$ and $d_2$, respectively. The corresponding section lengths are determined from the quarter-wave conditions $d_1 = \lambda_0/(4n_1)$ and $d_2 = n_1 d_1/n_2$. For $\lambda_0$ = 1548.35 nm, $n_1$ = 1.44, and $n_2$ = 1.4435, these relations give $d_1$ = 268.81 nm and $d_2$ = 268.16 nm. The physical length of each FP cavity is defined as $l_c = 21001 d_2$, corresponding to $l_c$ = 5.632 mm for the adopted design. The ring is coupled through ideal directional couplers, with coupling assumed to occur at a single point and to be lossless, consistent with the original model.

The narrow transmission resonance of the dual-cavity BGRR can be continuously tuned within the stopband defined by the Bragg mirrors by varying the refractive indices of the two FP cavities. For the uniform Bragg gratings considered in the present design, the stopband extends over approximately 300 GHz and defines the spectral region within which the narrow transmission resonance may be tuned. Since the desired millimeter-wave or sub-terahertz signal is generated by heterodyne photodetection of the filtered comb line and the optical carrier of the 1550-nm laser, the filter is designed with a center wavelength of approximately 1548.35 nm, corresponding to a frequency offset of approximately 206 GHz from the optical carrier. This choice places the filter approximately midway between the target 183- and 229-GHz frequency offsets and allows both corresponding comb components to lie well within the Bragg stopband, enabling either component to be selected through thermo-optic tuning.

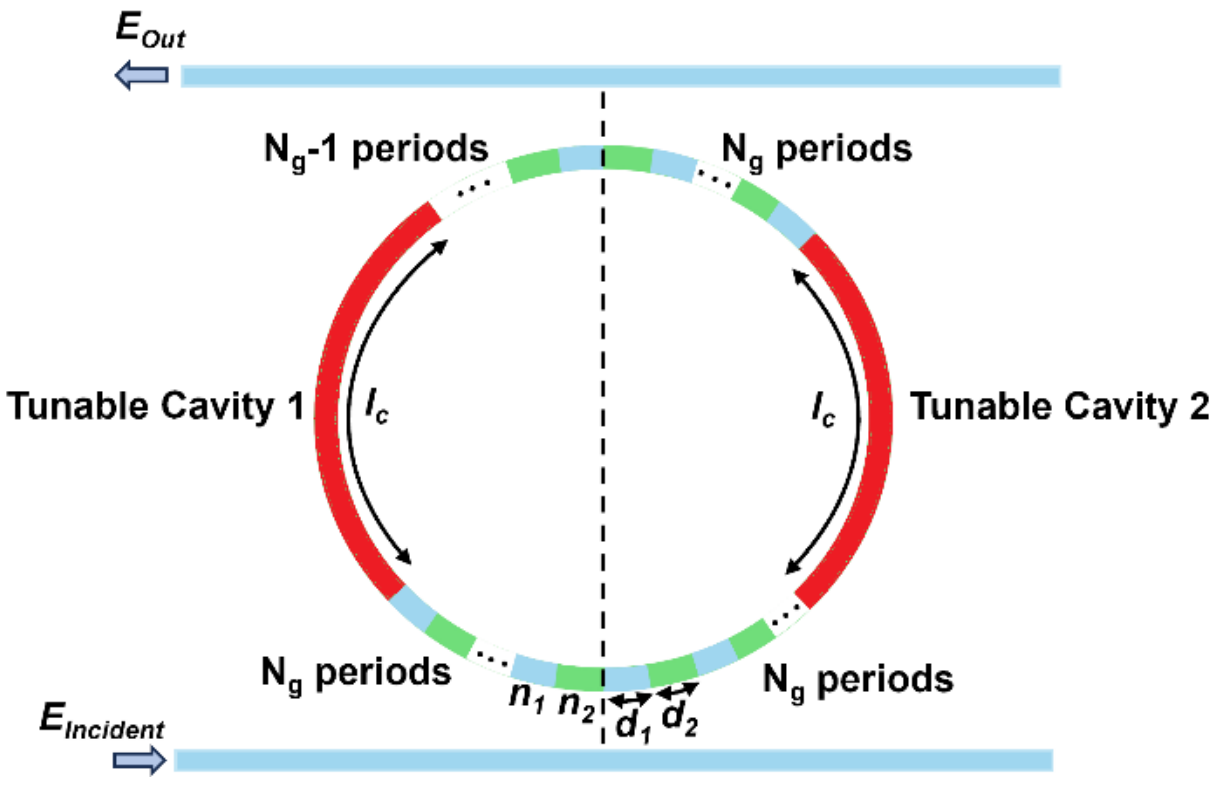

**Figure 2**. Block diagram of the dual-cavity Bragg grating ring resonator (BGRR) optical filter proposed in [10]. The two Fabry–Perot (FP) cavities, highlighted in red, correspond to the regions subjected to thermo-optic tuning. Adapted from [10].

The main parameters of the dual-cavity BGRR wavelength-selective filter used in the transfer-matrix model are summarized in Table 2.

Table 2. Design parameters of the dual-cavity Bragg grating ring resonator wavelength-selective filter.

| **Refractive index of low-index grating section ($n_1$)** | **1.44** |
|---|---|
| **Refractive index of high-index grating section ($n_2$)** | 1.4435 |
| **Number of grating periods per mirror ($N_g$)** | 1000 |
| **Central wavelength ($\lambda_0$)** | 1548.35 nm |
| **Physical length of the low-index grating section ($d_1$)** | 268.81 nm |
| **Physical length of the high-index grating section ($d_2$)** | 268.16 nm |
| **Amplitude loss coefficient per grating section di ($\alpha_i$)** | $10^{-10}$ |
| **FP cavity physical length ($l_c$)** | 5.632 mm |
| **Upper and lower ring-fiber self-coupling coefficients** | 0.96 |

To illustrate the wavelength-selection process for the two LO frequencies investigated in this work, Fig. 3 shows the theoretically calculated transmission responses of the 10-GHz coarse optical band-pass filter and the tunable dual-cavity BGRR, together with their combined response, for selection of the 183- and 229-GHz comb components. The corresponding comb-line positions are indicated schematically.

More generally, the proposed architecture could potentially be used to generate several LO frequencies relevant to the Earth-observation instruments reported in Table 1, including 89, 118, 166, 183, 229, 243, and 325 GHz, within the tuning range supported by the considered wavelength-selective filtering stage. Extension to higher LO frequencies, such as 448 GHz, would require a broader-band filtering implementation, for example based on chirped fiber Bragg gratings [14] or an alternative wavelength-selective architecture.

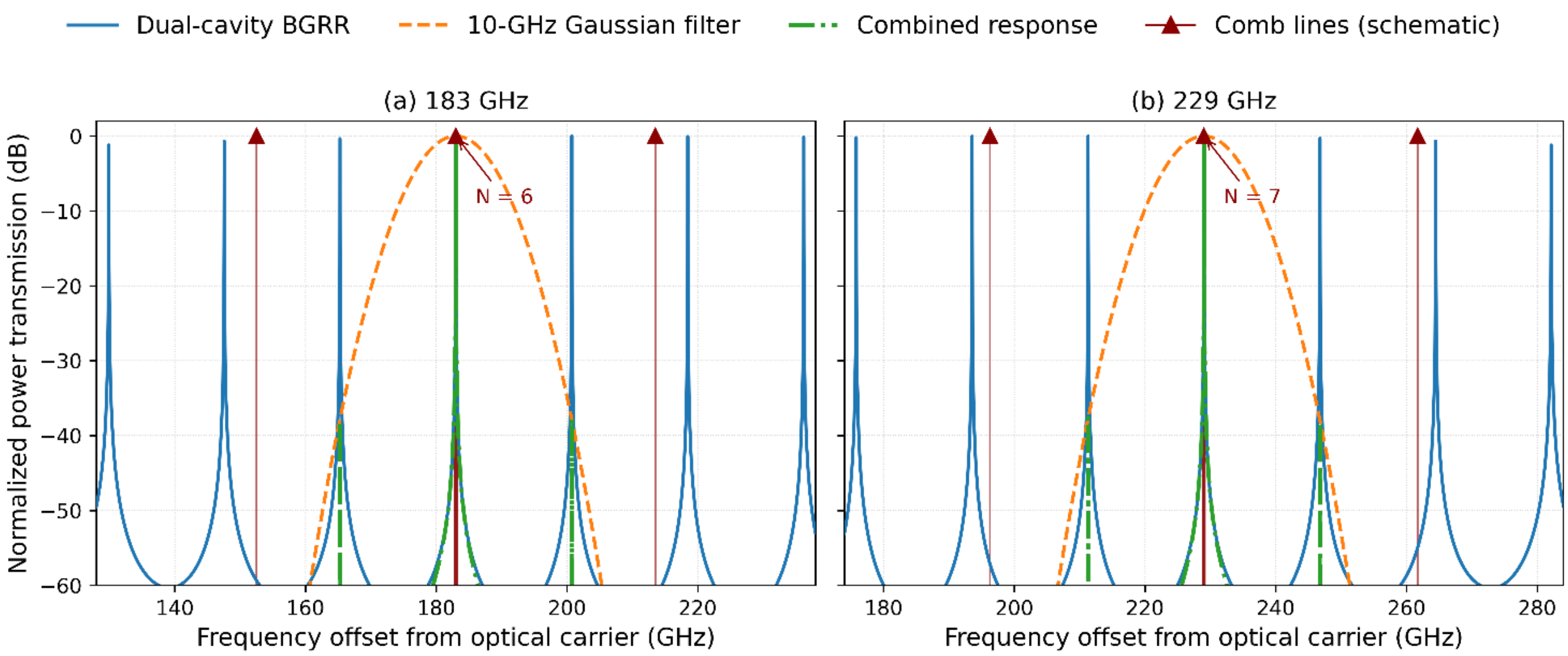

Figure 3. Theoretically calculated transmission responses of the 10-GHz coarse optical band-pass filter and the dual-cavity BGRR, together with their combined response, for selection of the (a) 183-GHz and (b) 229-GHz comb components. The corresponding comb-line frequencies are marked for reference.

## 3. SIMULATION SETUP

The proposed architecture was implemented and evaluated in VPIphotonics, with the system-level simulation model shown in Fig. 4. The self-oscillating loop was simulated in time-domain Sample Mode to capture the dynamic build-up and steady-state behavior of the OEO. A total of 30 simulation runs were performed for each investigated case. No appreciable differences were observed between the spectra obtained after 20 and 30 runs, indicating that the oscillation had effectively reached a steady-state condition. A simulation time window of 409.6 ns provides an intrinsic spectral resolution of 2.44 MHz. Sampling rates of 640 GHz and 1.28 THz were employed for the 183 GHz and 229 GHz generation cases, respectively.

The custom wavelength-selective optical filter presented at the end of the previous section was implemented through Python co-simulation using the complex transfer function obtained from the TMM of the resonant photonic structure. Since the VPI optical filtering routine operates on block signals, temporary Sample-to-Block and Block-to-Sample conversion was employed around the co-simulation stage, while the remaining system was maintained in Sample Mode.

The numerical simulations presented in this work focus on OEO operating frequencies around 30 GHz. The 183 GHz and 229 GHz LO-generation cases are investigated as representative examples of the general frequency-generation concept described in the previous section.

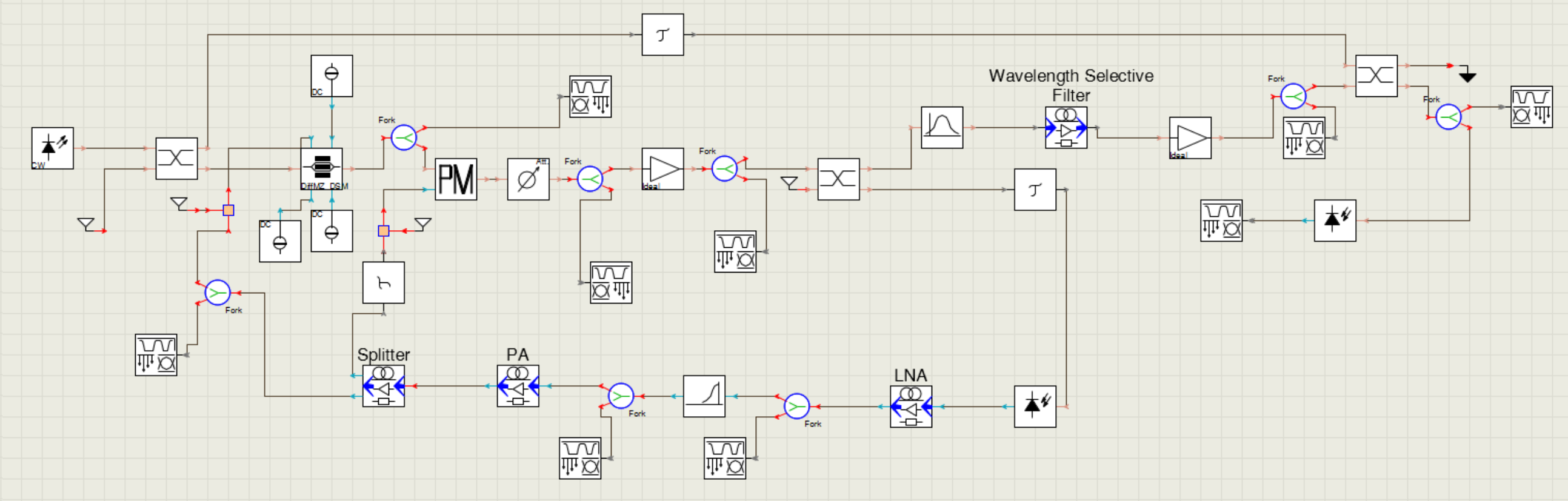

Figure 4. VPIphotonics system-level simulation model of the proposed self-oscillating photonic frequency-comb architecture.

The main parameters used in the VPIphotonics system-level simulations are summarized in Table 3. The selected values define the numerical simulation conditions, the OEO operating range, and the main parameters of the electro-optic frequency-generation, amplification and filtering stages. The simulation settings were chosen to provide sufficient spectral resolution over the frequency range of interest, while the OEO and filtering parameters were selected according to the targeted local-oscillator generation scenarios considered in this work.

A 7.2 ps delay is introduced between the RF drive paths of the MZM and PM, corresponding to a relative phase shift of approximately 79° at 30.5 GHz and 85° at 32.7 GHz.

The numerical model assumes optical and RF power levels appropriate for assessing the proposed frequency-generation principle. The component parameters were selected, where possible, to be representative of realistic device performance and experimentally achievable operating conditions. Nevertheless, an experimental implementation would require further consideration of insertion losses, amplification requirements, device power handling, photodetector bandwidth and saturation, as well as other practical non-idealities.

Table 3. Main parameters of the VPIphotonics system-level simulation model.

| Parameter | Value / Range |
|---|---|
| **Simulation platform** | VPIphotonics |
| **Main simulation mode** | Sample Mode |
| **Simulation time window** | 409.6 ns |
| **Frequency resolution** | 2.44 MHz |
| **Sampling rate** | 640 GHz for 183 GHz generation and 1.28 THz for 229 GHz generation |
| **Number of simulation runs** | 30 |
| **OEO operating-frequency range** | Around 30 GHz |
| **Representative LO frequencies** | 183 GHz, 229 GHz |
| **Electrical band-pass filter center frequency** | Tuned around the required OEO frequency |
| **Electrical band-pass filter bandwidth** | 10 MHz |
| **Low-noise amplifier gain** | 28 dB |
| **Low-noise amplifier NF** | 3 dB |
| **Power-amplifier gain** | 30 dB |
| **RF delay between MZM and PM drive paths** | 7.2 ps |
| **Laser Power** | 13 dBm |
| **Laser Emission Frequency** | 193.414 THz |
| **Loop delay** | 0.05 µs |
| **MZM half-wave voltage, ($V\pi$)** | 6 V |
| **MZM Insertion Loss** | 5 dB |
| **MZM DC bias** | 5 V |
| **PM half-wave voltage, ($V\pi$)** | 6 V |
| **PM insertion loss** | 3 dB |
| **Loop Optical Amplifier Gain** | 28 dB |
| **Loop Optical Amplifier NF** | 4 dB |
| **Optical Filter Bandwidth (coarse filtering)** | 10 GHz |
| **Heterodyne stage Optical Amplifier Gain** | 30 dB |
| **Heterodyne stage Optical Amplifier NF** | 4 dB |
| **Wavelength-selective filter** | Dual-cavity Bragg grating ring resonator |
| **Wavelength-selective filter model** | Transfer-matrix model |
| **Wavelength-selective filter Bandwidth** | 3 MHz |
| **WSF implementation** | Python co-simulation |
| **Photodetector responsivity** | 0.7 A/W |

## 4. SIMULATION RESULTS

The proposed self-oscillating photonic frequency-comb architecture was evaluated through system-level simulations in VPIphotonics. The analysis focuses on representative local-oscillator frequencies that can be generated using OEO operating frequencies around 30 GHz, in accordance with the frequency-multiplication concept introduced in the previous section.

For both 183 GHz and 229 GHz cases, the evolution of the OEO-generated microwave signal, the resulting optical frequency comb, the selection of the required comb component, and the final electrical signal obtained after photodetection were examined. Particular attention was given to the spectral purity of the generated LO signal and to the suppression of unwanted frequency components. The results presented below therefore include both close-in spectra around the generated carrier and wide-span spectra used to assess the presence of spurious components over the simulated frequency range.

**183 GHz Generation**

For the 183 GHz case, the OEO operates at a frequency close to 30.5 GHz and the sixth-order optical comb component is selected. After optical filtering, recombination with the optical carrier, and photodetection, a dominant electrical component is generated at the targeted 183 GHz frequency.

The simulated electrical output spectrum of the OEO is shown in Fig. 5. A dominant oscillation component is obtained around 30.5 GHz, within the passband of the electrical band-pass filter. This oscillation frequency determines the spacing of the subsequently generated optical frequency comb and therefore sets the frequency-multiplication factor required for the 183 GHz LO generation case. The spectrum confirms that the OEO operates with a clearly dominant microwave component over the displayed frequency range.

The optical spectrum at the output of the cascaded MZM–PM stage is shown in Fig. 6. A broad frequency comb is generated with a line spacing determined by the OEO oscillation frequency. For the 183 GHz generation case, the sixth-order comb component is selected for subsequent beating with the optical carrier. Although the generated comb exhibits a non-uniform spectral envelope, comb flatness is not a primary requirement for the proposed LO-generation approach. The relevant condition is the availability of the required comb order with sufficient optical power for spectral selection and subsequent photodetection.

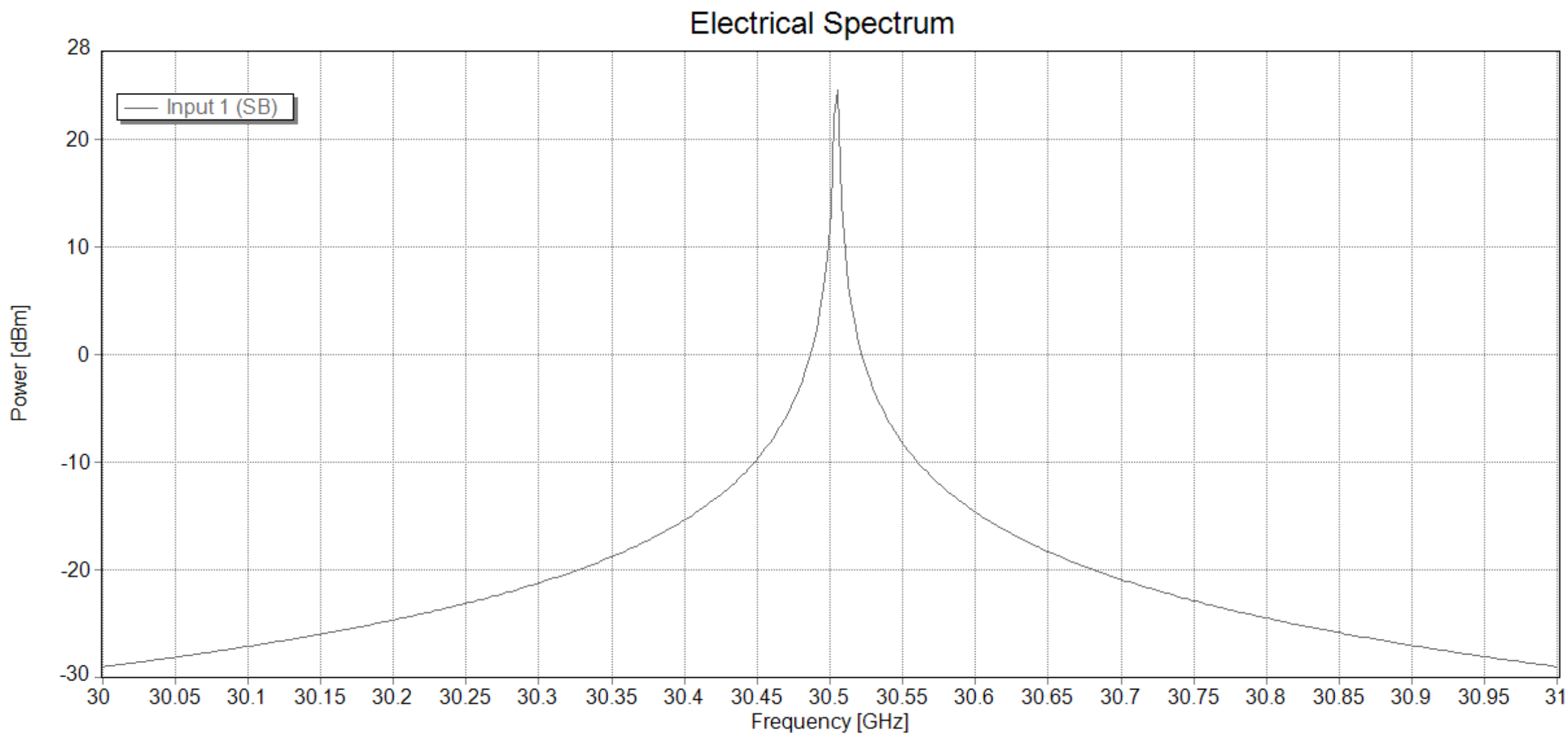

Figure 5. Simulated electrical output spectrum of the OEO for the 183 GHz LO generation case. The dominant oscillation component around 30.5 GHz determines the spacing of the generated optical frequency comb. The intrinsic spectral resolution is 2.44 MHz.

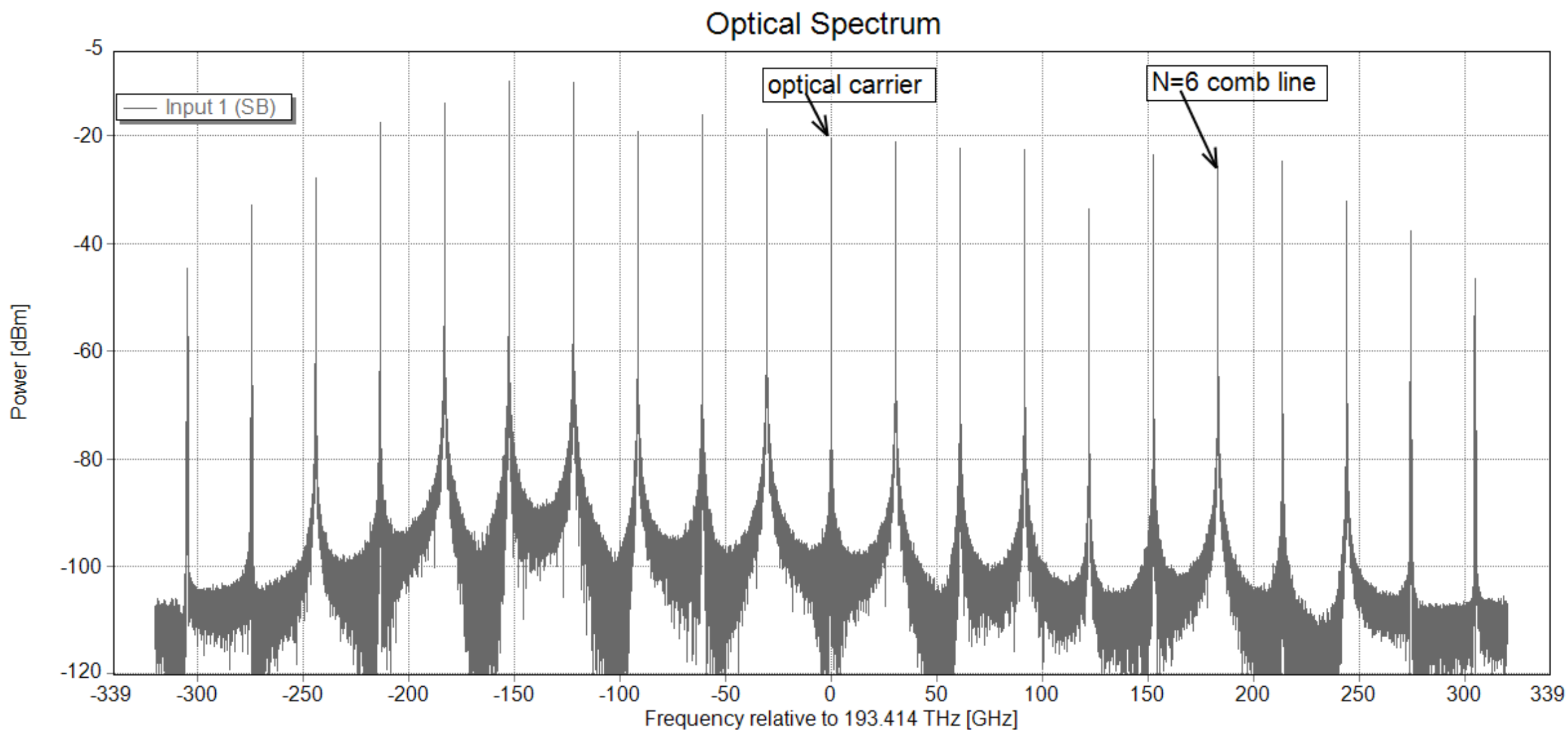

Figure 6. Simulated optical spectrum at the output of the cascaded MZM–PM comb-generation stage for the 183 GHz LO generation case. The comb lines are spaced by the OEO oscillation frequency, with the sixth-order component selected for subsequent optical filtering and photodetection.

A close-in view of the generated 183 GHz signal is shown in Fig. 7. The spectrum is dominated by a single carrier, while the surrounding spectral pedestal remains substantially below the carrier level. No strong adjacent discrete spectral components are observed within the displayed frequency range.

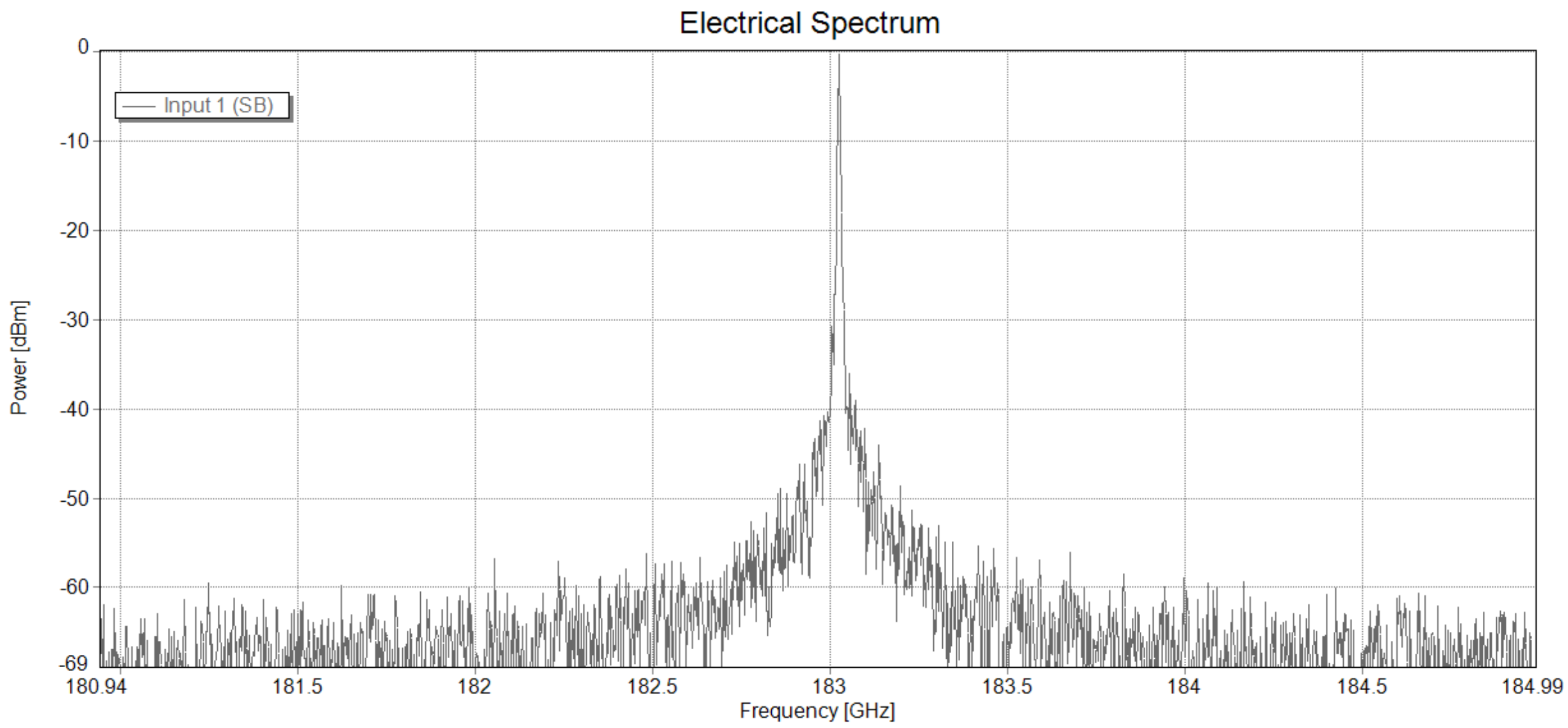

Figure 7. Close-in electrical spectrum of the generated 183 GHz signal. The intrinsic spectral resolution is 2.44 MHz.

The wide-span electrical spectrum is shown in Fig. 8. The desired 183 GHz component clearly dominates the output spectrum, while the remaining spectral components are strongly suppressed throughout the investigated frequency range. Using horizontal markers to compare the level of the desired carrier with the highest unwanted spectral component, a spurious suppression of 55 dB is obtained. Together with the generated output power of −0.16 dBm, this result demonstrates that the proposed architecture can generate the targeted sub-THz LO signal while maintaining strong suppression of unwanted frequency products.

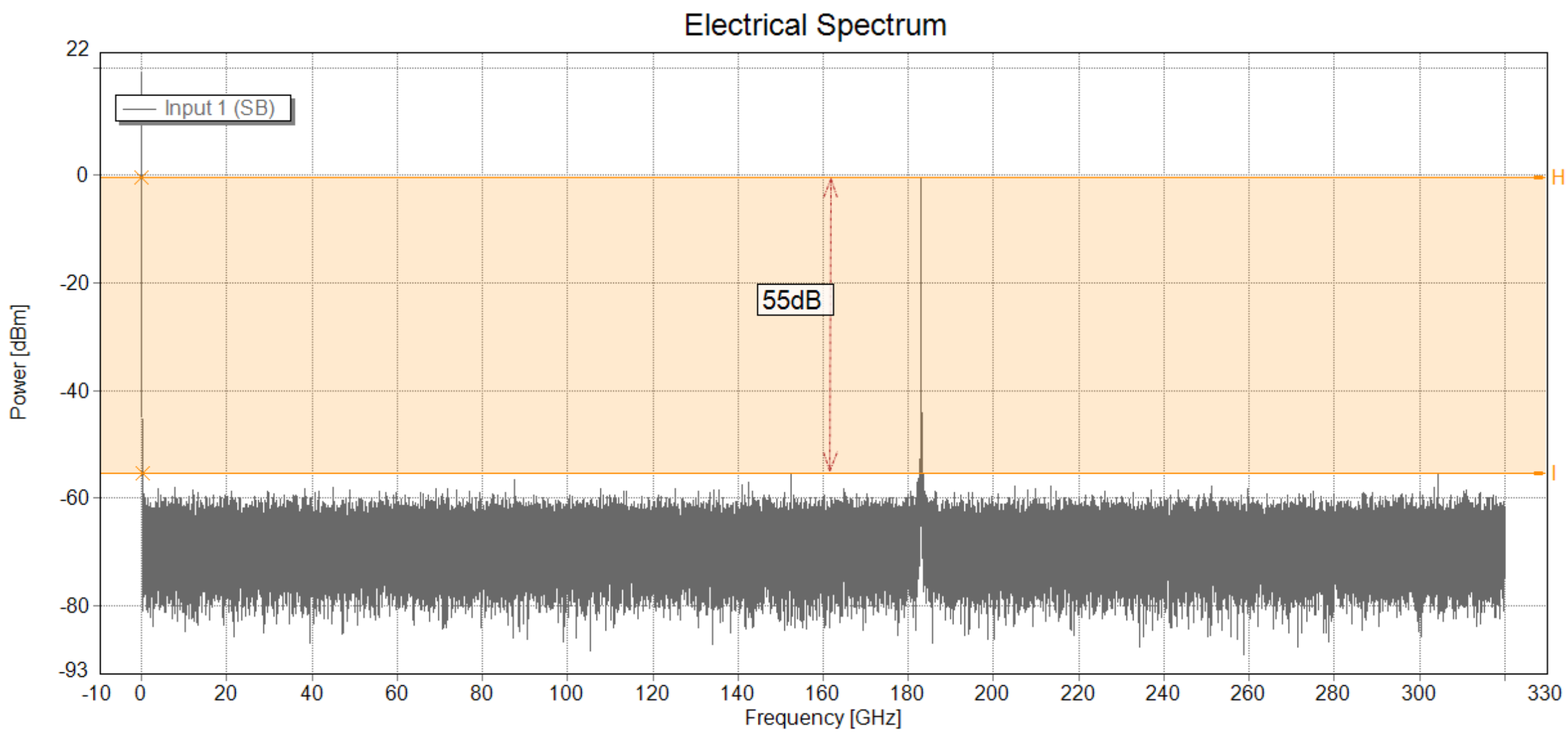

Figure 8. Simulated wide-span electrical spectrum of the generated 183 GHz signal, illustrating suppression of unwanted spurious components. The intrinsic spectral resolution is 2.44 MHz.

**229 GHz Generation**

For the 229 GHz LO generation case, the OEO operates at approximately 32.7 GHz. As shown in Fig. 9, a dominant oscillation component is established around 32.7 GHz, which determines the spacing of the subsequently generated optical frequency comb.

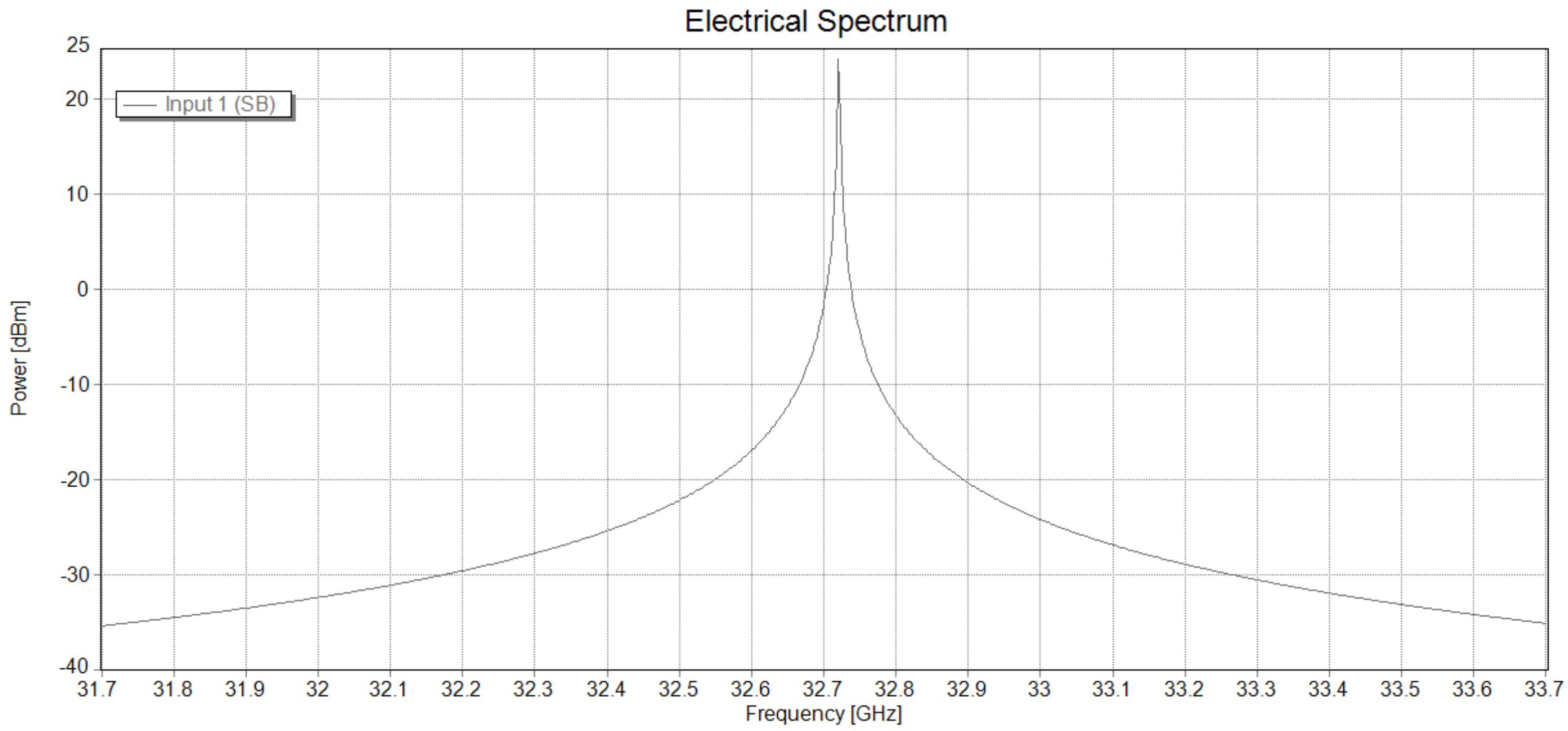

Figure 9. Simulated electrical output spectrum of the OEO for the 229 GHz LO generation case. The dominant oscillation component around 32.7 GHz determines the spacing of the generated optical frequency comb. The intrinsic spectral resolution is 2.44 MHz.

The optical spectrum at the output of the cascaded MZM–PM comb-generation stage is shown in Fig. 10. Multiple comb components are generated with a frequency spacing defined by the OEO oscillation frequency. The comb lines up to N = 16 appear in Fig. 10 with observable powers. For this case, the seventh-order comb component is selected by the wavelength-selective optical filter and recombined with the optical carrier. Photodetection of the two optical components produces the required electrical signal in the vicinity of 229 GHz.

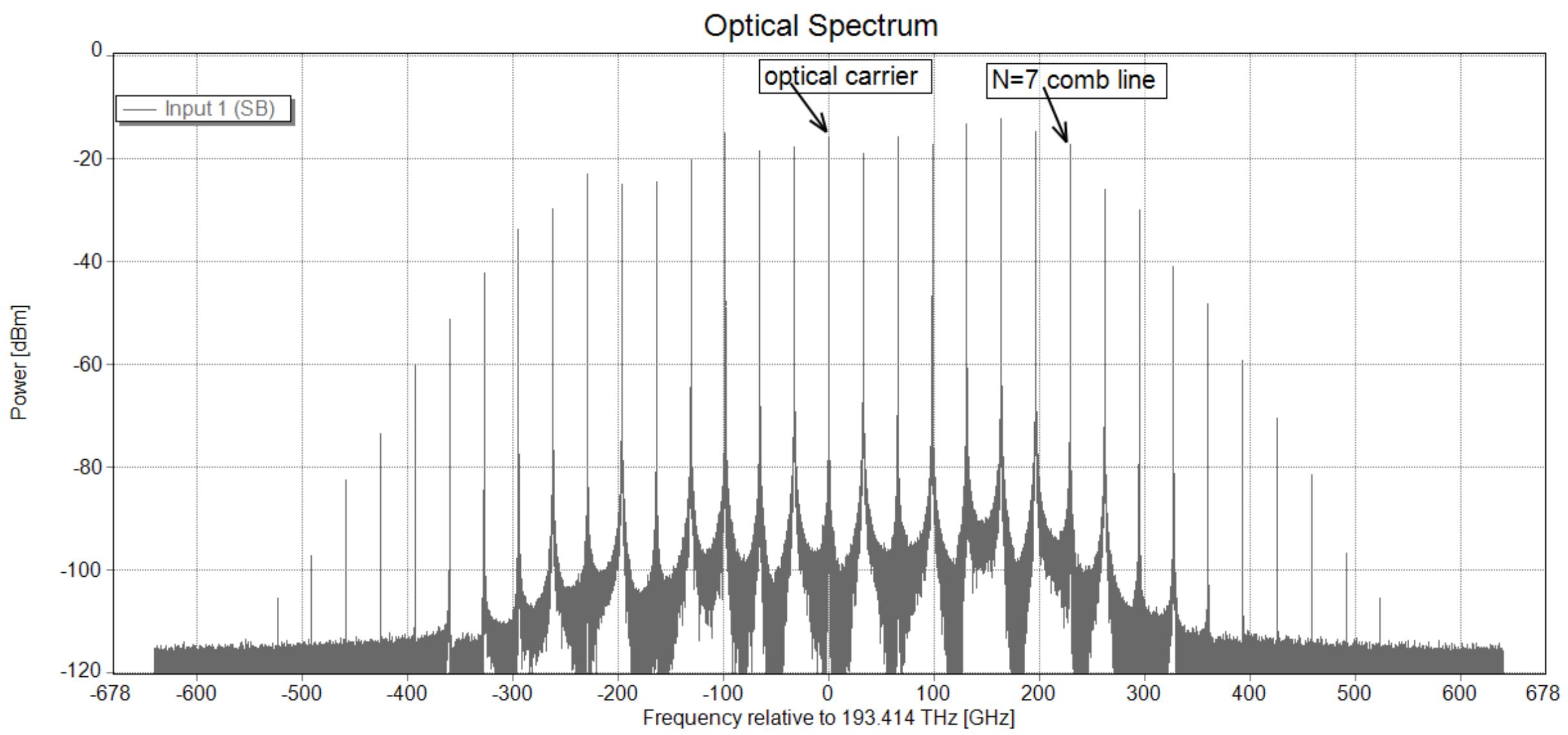

Figure 10. Simulated optical spectrum at the output of the cascaded MZM–PM comb-generation stage for the 229 GHz LO generation case. The comb lines are spaced by the OEO oscillation frequency, with the seventh-order component selected for subsequent optical filtering and photodetection.

Figure 11 presents a close-in view of the generated 229 GHz signal. The spectrum is dominated by the desired carrier, while the surrounding spectral components remain well below the main tone. The output power of the 229 GHz signal is +5.77 dBm.

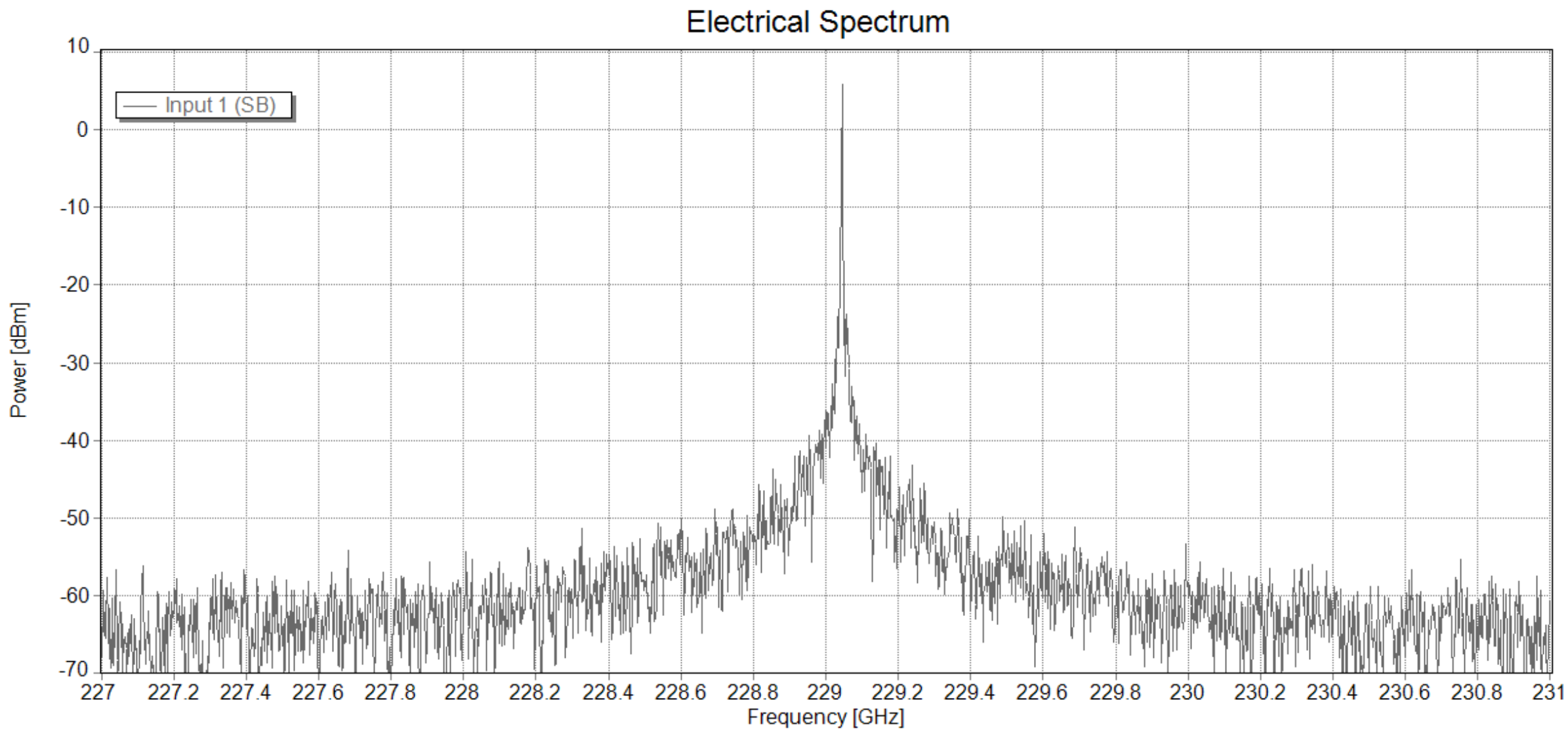

Figure 11. Close-in electrical spectrum of the generated 229 GHz signal. The intrinsic spectral resolution is 2.44 MHz.

The wide-span electrical spectrum is shown in Fig. 12. The desired 229 GHz component clearly dominates the output over the investigated frequency range, while unwanted frequency components are strongly suppressed by at least 59 dB. Together with the close-in spectrum, these results demonstrate the generation and spectral selection of the 229 GHz LO signal using the same self-oscillating photonic frequency-comb architecture employed for the 183 GHz case. The higher suppression achieved in the 229 GHz case can be attributed, at least in part, to the higher optical power of the seventh-order comb line used for 229 GHz generation compared with that of the sixth-order comb line employed in the 183 GHz case. Further improvement in the generated LO power and spectral suppression may be achieved by optimizing the electrical drive powers applied to the modulators together with the relative RF phase shift between them, since these parameters directly affect the distribution of optical power among the generated comb lines

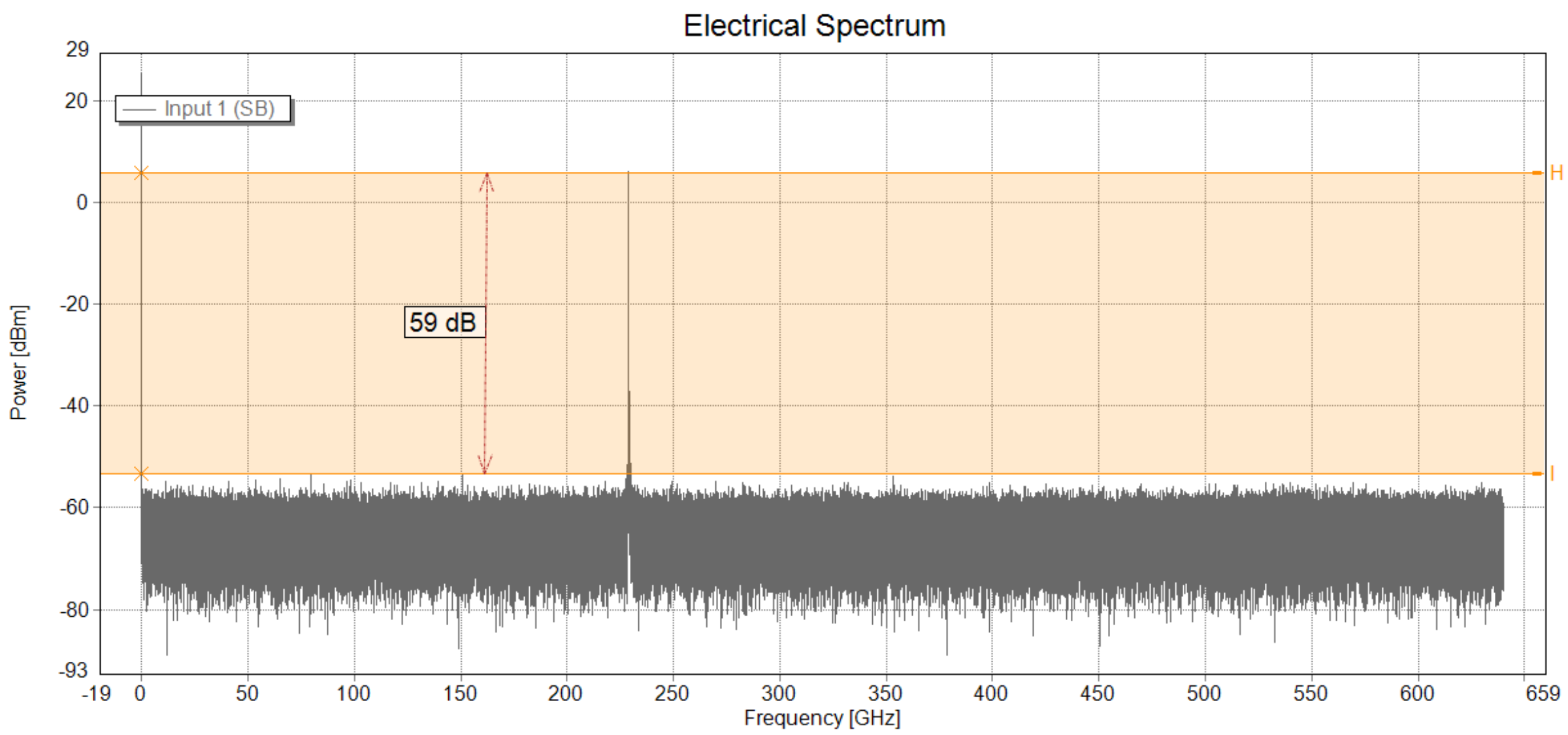

Figure 12. Simulated wide-span electrical spectrum of the generated 229 GHz signal, illustrating suppression of unwanted spurious components. The intrinsic spectral resolution is 2.44 MHz.

## 5. CONCLUSIONS

This work has proposed and numerically investigated a reconfigurable self-oscillating frequency-comb architecture for millimeter-wave and sub-terahertz local-oscillator generation in Earth-observation systems. The approach combines an optoelectronic oscillator with cascaded Mach–Zehnder and phase modulation to generate an optical frequency comb, whose spacing is determined by the OEO oscillation frequency. The required higher-order comb component is isolated using a conventional optical bandpass filter followed by a thermo-optically tunable dual-cavity Bragg-grating ring resonator for narrowband selection and is then heterodyned with an optical reference derived from the same laser. This enables the OEO and electro-optic modulators to operate at microwave frequencies well below the targeted LO frequency.

System-level VPIphotonics simulations were performed for 183- and 229-GHz generation using OEO frequencies of approximately 30.5 and 32.7 GHz, respectively. For the 183-GHz case, selection of the sixth-order comb component produced the targeted signal with a simulated output power of −0.16 dBm and approximately 55-dB suppression of the strongest unwanted spectral component. For the 229-GHz case, selection of the seventh-order comb component produced an output power of 5.77 dBm, while unwanted frequency components were suppressed by at least 59 dB. The higher output power and suppression in the latter case are partly associated with the greater optical power of the selected seventh-order comb line. Further improvement in LO power and spectral suppression could be achieved by optimizing the modulator RF drive powers and their relative phase shift.

The analysis of the accessible LO frequencies indicates that several frequencies relevant to MWS, MWI, and ICI could potentially be addressed using the same architecture through appropriate selection of the OEO frequency and comb order, reducing the need for dedicated oscillator and frequency-multiplication chains for individual bands. The proposed system is conceived as a hybrid photonic–electronic platform, with selected functions suitable for photonic and potentially electronic integration. Experimental validation and further assessment of practical optical and RF power budgets, component limitations, and integration constraints would be required to establish the feasibility of the concept as a compact and reconfigurable multi-frequency LO source for spaceborne Earth-observation instrumentation.

## ACKNOWLEDGEMENTS

This work was co-funded by the European Union and the Republic of Cyprus through the Research and Innovation Foundation under the project PICSOC (Grant ID: EXCELLENCE/0524/0507). The authors would also like to thank Enrico Lia of the European Space Agency (ESA) for providing reference material that supported the preparation of Table 1.